\documentstyle[eqsecnum,aps,pre,epsfig]{revtex}
\DeclareFontShape{OML}{cmm}{m}{b}{%
   <-> cmmib10}{}
\DeclareMathAlphabet{\mathbf}{OML}{cmm}{m}{b}
\DeclareSymbolFont{boldletters}{OML}{cmm}{m}{b}
\DeclareMathSymbol{\balpha}{\mathord}{boldletters}{11}
\DeclareMathSymbol{\bbeta}{\mathord}{boldletters}{12}
\DeclareMathSymbol{\bgamma}{\mathord}{boldletters}{13}
\DeclareMathSymbol{\bomega}{\mathord}{boldletters}{33}
\DeclareMathSymbol{\bsigma}{\mathord}{boldletters}{27}
\DeclareMathSymbol{\btau}{\mathord}{boldletters}{28}

\def\Tup#1#2#3{\setlength{\unitlength}{.5mm}
\makebox{
\begin{picture}(16,11)(-2,3.15)
\thicklines
\put(0,0){\line(1,0){10}}
\put(0,0){\line(3,5){5}}
\put(10,0){\line(-3,5){5}}
\put(13,0){\makebox(0,0)[c]{\tiny #3}}
\put(-3,0){\makebox(0,0)[c]{\tiny #2}}
\put(5,11.5){\makebox(0,0)[c]{\tiny #1}}
\end{picture}}}

\def\case#1#2{\textstyle{#1\over#2}\displaystyle}
\def\i{{\rm i}}
\def\d{{\rm d}}
\def\e{{\rm e}}

\begin{document}
\title{Phase diagram of the $su(8)$ quantum spin tube}
\author{J. de Gier, M.~T. Batchelor and M. Maslen}
\address{Department of Mathematics, School of Mathematical Sciences,\\
Australian National University, Canberra ACT 0200, Australia}
\date{\today}
\draft
\maketitle
\begin{abstract}
We calculate the phase diagram of an integrable anisotropic
3-leg quantum spin tube connected to the $su(8)$ algebra. We find
several quantum phase transitions for antiferromagnetic rung
couplings. Their locations are calculated exactly from the Bethe
Ansatz solution and we discuss the nature of each of the different
phases.
\end{abstract}

\section{Introduction}

The properties of $n$-leg quantum spin ladders have attracted
considerable recent attention from both theorists and 
experimentalists \cite{DR}.
The main reasons for this interest are that, like their one-dimensional
counterparts, ladders are amenable to approximation schemes such as 
bosonization, and to accurate numerical investigation by the DMRG
method.    
It is now well established that the $n$-leg
Heisenberg ladders have the rather striking property of being
massive for $n$ even and massless for $n$ odd.
A number of ladder-like compounds have been found.   
For example, the 2-leg ladders $SrCu_2O_3$ \cite{A} and 
$Cu_2(C_5H_{12}N_2)_2Cl_4$ \cite{C} exhibit a gap, while the
3-leg ladder $Sr_2Cu_3O_5$ \cite{A} is gapless.

Although the Heisenberg ladders are not solvable in the sense of the
Heisenberg spin chain, a number of solvable ladder models are 
known \cite{FR,AFW,BMb,W,BMa,F,FK,BGLM}.
One 2-leg ladder model with fixed rung, leg and diagonal interactions 
is equivalent to the XXZ chain at $\Delta=-\frac{5}{3}$ 
and is thus massive \cite{AFW,BMb}.   
Another 2-leg model has been constructed which is massless in the absence
of rung interactions with a transition to a massive phase at a critical
(non-zero) value of the rung coupling \cite{W}.
This is in contrast with the 2-leg Heisenberg spin ladder for which 
the transition to the massive phase is at zero rung coupling.   
In the integrable model the Heisenberg rung interactions appear as 
chemical potentials which break the underlying $su(4)$ symmetry.
The phase diagram of this model has been established by means of the
Bethe Ansatz solution \cite{W}.
The integrable model has been generalised to an arbitrary number of legs 
in which the underlying $su(2^n)$ symmetry is broken
by the Heisenberg rung interactions, which again appear as
chemical potentials in the Bethe Ansatz solution \cite{BMa}.
Integrable ladder models with Hubbard \cite{F} and $t-J$ \cite{FK} 
interactions have also been recently found.
Others have been constructed with $o(2^n)$ and $sp(2^n)$ 
symmetry \cite{BGLM}.
Some ladder models have also been constructed with dimer or
matrix product ground states \cite{mp}.

In this paper we investigate the thermodynamic properties of the
solvable 3-leg qnantum spin tube \cite{BMa}.

\section{The Model}
\label{se:model}
We consider an integrable 3-leg spin ladder, or spin tube for
periodic boundary conditions. The spins along each leg and each rung
have an isotropic Heisenberg interaction, with the introduction 
of many-body terms to retain integrability. As usual, it is
convenient to express the Hamiltonian in terms of the Pauli matrices,  
\begin{equation}
\sigma^x = \left( \begin{array}{@{}cc@{}} 0 & 1 \\ 1 & 0 \end{array}
\right), \quad \sigma^y = \left( \begin{array}{@{}cc@{}} 0 & -\i \\ \i &
0 \end{array} \right), \quad \sigma^z = \left( \begin{array}{@{}cc@{}}
1 & 0 \\ 0 & -1 \end{array} \right). 
\end{equation}
The Hamiltonian of our model is
\begin{equation}
H = \sum_{i=1}^L h^{\rm leg}_{i,i+1} + \sum_{i=1}^L h^{\rm
rung}_i,
\label{eq:Ham}
\end{equation}
where, taking the numbering on each elementary triangular plaquette
to be $\Tup{1}{2}{3}$,
\begin{eqnarray}
h^{\rm leg}_{i,j} &=& \case{1}{8} \prod_{l=1}^3 \left( 1+
\bsigma_i^{(l)} \cdot \bsigma_{j}^{(l)}\right), \label{eq:hleg}\\
h^{\rm rung}_i &=& \sum_{l=1}^3 \case{1}{2} J_l \left(
\bsigma_i^{(l)} \cdot \bsigma_i^{(l+1)} -1 \right).\label{eq:hrung}
\end{eqnarray}
The operators $(\sigma^x)_i^{(l)}$, $(\sigma^y)_i^{(l)}$ and
$(\sigma^z)_i^{(l)}$ act as the corresponding Pauli matrices on the
$(i,l)$th factor in the Hilbert space,  
\begin{equation}
{\mathcal H} = \bigotimes_{i=1}^L V_i,\quad V_i = \bigotimes_{l=1}^3
{\mathbf C}^2. 
\end{equation}
We take periodic boundary conditions along and also across the
ladder. Now we note that $h^{\rm leg}_{i,j}$ is just the permutator on
the space $V_i \otimes V_j$ and thus that (\ref{eq:hleg}) is simply
the integrable isotropic permutation Hamiltonian corresponding to 
the $su(8)$ algebra \cite{S}. Since 
$[h^{\rm leg}_{i,j}, h^{\rm rung}_i + h^{\rm rung}_{i+1}] =0$, 
$H$ defined in (\ref{eq:Ham}) is also integrable. We
set $J_1=J_2=J$ and $J_3=J'$, so that we can go from the isotropic 
tube, $J'=J$, to the ladder, $J'=0$. This model thus generalizes 
that given in \cite{BMa} to anisotropic rung couplings.  

It is convenient to change to the basis where the square and the
$z$-component of the total spin of a given triangle, ${\mathbf S}=
\bsigma^{(1)} + \bsigma^{(2)} + \bsigma^{(3)}$ are diagonal. It
follows that the eight states on a given triangle fall into a
spin-$\frac{3}{2}$ quadruplet and two spin-$\frac{1}{2}$ doublets. 
This basis, in terms of the $(\sigma^z)^{(l)}$ eigenvalues, is,
\begin{eqnarray} 
|0\rangle &=& |\case{1}{2};\case{1}{2}\rangle_1 = \case{1}{\sqrt{6}}
\left( 2\Tup{$+$}{$-$}{$+$} - \Tup{$+$}{$+$}{$-$} -
\Tup{$-$}{$+$}{$+$} \right )\nonumber\\ 
|1\rangle &=& |\case{1}{2};-\case{1}{2}\rangle_1 = \case{1}{\sqrt{6}}
\left( 2\Tup{$-$}{$+$}{$-$} - \Tup{$-$}{$-$}{$+$} -
\Tup{$+$}{$-$}{$-$} \right )\nonumber\\
|2\rangle &=& |\case{1}{2};\case{1}{2}\rangle_2 = \case{1}{\sqrt{2}}
\left( \Tup{$+$}{$+$}{$-$} - \Tup{$-$}{$+$}{$+$} \right )\nonumber\\ 
|3\rangle &=& |\case{1}{2};-\case{1}{2}\rangle_2 = \case{1}{\sqrt{2}}
\left( \Tup{$-$}{$-$}{$+$} - \Tup{$+$}{$-$}{$-$} \right ) \nonumber\\
\label{eq:basis}\\
|4\rangle &=& |\case{3}{2};\case{3}{2}\rangle =
\vphantom{\case{1}{\sqrt{3}}} \Tup{$+$}{$+$}{$+$} \nonumber\\ 
|5\rangle &=& |\case{3}{2};\case{1}{2}\rangle = \case{1}{\sqrt{3}}
\left( \Tup{$-$}{$+$}{$+$} + \Tup{$+$}{$-$}{$+$} + \Tup{$+$}{$+$}{$-$}
\right ) \nonumber\\ 
|6\rangle &=& |\case{3}{2};-\case{1}{2}\rangle = \case{1}{\sqrt{3}}
\left( \Tup{$+$}{$-$}{$-$} + \Tup{$-$}{$+$}{$-$} + \Tup{$-$}{$-$}{$+$}
\right) \nonumber\\ 
|7\rangle &=& |\case{3}{2};-\case{3}{2}\rangle = \Tup{$-$}{$-$}{$-$}
\nonumber 
\end{eqnarray}
It is emphasized that the Hamiltonian (\ref{eq:hleg}) has the same
form on this new basis. The rung Hamiltonian (\ref{eq:hrung}) now
becomes diagonal and is given by $h^{\rm rung}={\rm diag}
\{-3J,-3J,-2J'-J,-2J'-J,0,0,0,0\}$. 

As claimed above, $H$ can be diagonalized using the Bethe Ansatz. The
Bethe Ansatz equations are well known \cite{S}, and given by
\begin{eqnarray}
\left( \frac{\lambda_j^{(1)} - \frac{\i}{2}}{\lambda_j^{(1)} +
\frac{\i}{2}} \right)^L &=& \prod_{k \neq j}^{M_1}
\frac{\lambda_j^{(1)} - \lambda_k^{(1)} - \i}{\lambda_j^{(1)} -
\lambda_k^{(1)} + \i} \prod_{k=1}^{M_2}
\frac{\lambda_j^{(1)} - \lambda_k^{(2)} + \frac{\i}{2}}{\lambda_j^{(1)} -
\lambda_k^{(2)} - \frac{\i}{2}}, \nonumber\\
\\\label{eq:BAE}
\prod_{k \neq j}^{M_r}
\frac{\lambda_j^{(r)} - \lambda_k^{(r)} - \i}{\lambda_j^{(r)} -
\lambda_k^{(r)} + \i} &=& \prod_{k=1}^{M_{r-1}}
\frac{\lambda_j^{(r)} - \lambda_k^{(r-1)} - \frac{\i}{2}}{\lambda_j^{(r)} -
\lambda_k^{(r-1)} + \frac{\i}{2}} \prod_{k=1}^{M_{r+1}}
\frac{\lambda_j^{(r)} - \lambda_k^{(r+1)} - \frac{\i}{2}}{\lambda_j^{(r)} -
\lambda_k^{(r+1)} + \frac{\i}{2}},\nonumber     
\end{eqnarray}\
where $j=1,\ldots,M_r$ with $r=2,\ldots,7$ and $M_8=0$.
The eigenenergies of $\sum_{i=1}^L h^{\rm leg}_{i,i+1}$ are given by
\begin{equation}
E^{\rm leg} = -\sum_{j=1}^{M_1} \frac{1}{(\lambda_j^{(1)})^2 + \frac{1}{4}}.
\end{equation}
It is important to note that since the permutator (\ref{eq:hleg}) is
invariant under any ordering of the states (\ref{eq:basis}), this
result may be obtained using any choice of reference state (or
pseudo-vacuum) $|\Omega\rangle$ and any assignment of Bethe Ansatz
pseudo particles. For each choice however, one has to re-interpret the
numbers $M_r$ in terms of those corresponding to the ordering chosen
in (\ref{eq:basis}). The rung Hamiltonian does alter with the choice
of ordering, but the change is just a rearrangement of its eigenvalues
along the diagonal. We use this property to our advantage by doing
calculations with that choice of ordering for which the Bethe Ansatz
reference state is closest to the true ground state of the
system. Below we list the six different possibilities corresponding to
the choice of ordering of the two doublets and the quadruplet.

\begin{itemize}
\item[1.] $\{|0\rangle, |1\rangle, |2\rangle, |3\rangle, |4\rangle,
|5\rangle, |6\rangle, |7\rangle \},\qquad E = E^{\rm leg} +2(J-J')
M_2 + (2J'+J) M_4$.

The Bethe Ansatz reference state is
\[
|\Omega\rangle = \bigotimes_{i=1}^L |0\rangle_i = \bigotimes_{i=1}^L
|\case{1}{2};\case{1}{2}\rangle_1^i,
\]
and the assignment of pseudo particles is such that $M_k =
\sum_{r=k}^7 N_r$, where $N_r$ is the number of states $|r\rangle$
occuring in an excited state. In the cases below the reference state
is understood to be the tensor product of the first state listed.

\item[2.] $\{|2\rangle, |3\rangle, |0\rangle, |1\rangle, |4\rangle,
|5\rangle, |6\rangle, |7\rangle \},\qquad E = E^{\rm leg} + 2(J'-J)
M_2 + 3J M_4$.  

Now the pseudo particle assignment is given by $M_1 =
N_0+N_1+N_3+\sum_{r=4}^7 N_r$, $M_2 = N_0+N_1+\sum_{r=4}^7 N_r$, $M_3
= N_1+\sum_{r=4}^7 N_r$ and $M_k = \sum_{r=k}^7 N_r$ for $k \geq 4$.

\item[3.] $\{|2\rangle, |3\rangle, |4\rangle, |5\rangle, |6\rangle,
|7\rangle, |0\rangle, |1\rangle \},\qquad E = E^{\rm leg} +(2J'+J)M_2
- 3J M_6$,
  
with $M_k = \sum_{r=0}^1 N_r + \sum_{r=k+2}^7 N_r$ for
$1 \leq k \leq 5$, $M_6 = N_0+N_1$ and $M_7 = N_1$.

\item[4.] $\{|4\rangle, |5\rangle, |6\rangle, |7\rangle, |2\rangle,
|3\rangle, |0\rangle, |1\rangle \},\qquad E = E^{\rm leg} - (2J'+J)
M_4 + 2(J'-J) M_6$,

with $M_k = \sum_{r=0}^3 N_r + \sum_{r=k+4}^7 N_r$ for $k \leq 3$ and
$M_4 = N_0+N_1+N_2+N_3$, $M_5 = N_0+N_1+N_3$, $M_6 = N_0+N_1$ and $M_7
= N_1$.

\item[5.] $\{|4\rangle, |5\rangle, |6\rangle, |7\rangle, |0\rangle,
|1\rangle, |2\rangle, |3\rangle \},\qquad E = E^{\rm leg} - 3J M_4 +
2(J-J') M_6$,  

with $M_k = \sum_{r=0}^3 N_r + \sum_{r=k+4}^7 N_r$ for $k \leq 3$ and
$M_k = \sum_{r=k-4}^3 N_r$ for $k\geq 4$.

\item[6.] $\{|0\rangle, |1\rangle, |4\rangle, |5\rangle, |6\rangle,
|7\rangle, |2\rangle, |3\rangle \},\qquad E = E^{\rm leg} + 3J M_2 -
(2J'+J) M_6$,  

with $M_1=N_1+M_2$, $M_k = \sum_{r=2}^3 N_r + \sum_{r=k+2}^7 N_r$ for
$2 \leq k \leq 5$, $M_6 = N_2+N_3$ and $M_7 = N_3$.
\end{itemize}

Of course, all of these different choices for the reference state do not
change the physics, but it turns out that each of them is convenient
in some part of the phase diagram, as we will see in Section
\ref{se:phases}. Before calculating the complete phase diagram we
first consider two interesting lines.

\section{Special lines}

\subsection{The tube: $J'=J$}

Due to the extra symmetry in this case, the two doublets become
degenerate. Consider now the Bethe Ansatz where the reference state is
the spin up state of either of the spin-$\frac{1}{2}$
doublets. According to case 1. or 2. in Section \ref{se:model}, 
the energy is given by
\begin{equation}
E = -\sum_{i=1}^{M_1} \frac{1}{(\lambda_j^{(1)})^2 + \frac{1}{4}} + 3J M_4,
\end{equation}
where $M_4$ denotes the number of quadruplet excitations. It is
obvious that if $J$ is large and positive, no quadruplet state can
exist in the ground state. There are three Fermi seas and therefore
three gapless excitations. All flavours of the quadruplet excitations
are massive and their gap may be calculated by creating a
$\lambda^{(4)}$ mode and a $\lambda^{(3)}$ hole in the Bethe Ansatz
equation \cite{W} (see Appendix \ref{ap:gap}). The gap is
\begin{equation}
\Delta = 3J - \frac{1}{4}(\log 4 + \pi).
\end{equation}  
Thus we find a phase transiton from $J>J_{\rm c}^+ = \frac{1}{12}(\log 4
+ \pi)$ to $0 < J < J_{\rm c}^+$ where all quadruplet
flavours become massless and there are seven Fermi seas and no
massive excitations. Note that only at $J=0$, all Fermi seas are
completely filled.

Considering the case $J<0$, it is convenient to choose either case
4. or 5. of Section \ref{se:model}. The energy is given by
\begin{equation}
E = -\sum_{i=1}^{M_1} \frac{1}{(\lambda_j^{(1)})^2 + \frac{1}{4}} +
3|J| M_4, 
\end{equation}
where now $M_4$ denotes the number of excitations on the
spin-$\case{3}{2}$ ground state. These excitations are the four flavours
belonging to the two spin-$\case{1}{2}$ states and are all massive for
$J \ll 0$. We can use the same calculation as above to conclude that again
there is a phase transition at $J=J_{\rm c} = - J_{\rm c}^+$ where the
two spin-$\case{1}{2}$ doublet excitations become massless. 

\subsection{The ladder: $J'=0$}
\label{se:ladder}
As above, we start by considering this case for $J>0$. It is most
convenient to take the Bethe Ansatz as in case 1. of Section
\ref{se:model}. The energy is thus given by
\begin{equation}
E=-\sum_{i=1}^{M_1} \frac{1}{(\lambda_j^{(1)})^2 + \frac{1}{4}} + 2J
M_2 +J M_4.
\end{equation}
For $J \gg 0$ it is obvious that $M_2=0$ for the ground state. The
ground state is thus formed by the $\{|0\rangle,|1\rangle\}$ doublet
and there is one gapless excitation and six massive ones. As before we
can calculate the gap by introducing a $\lambda^{(2)}$ mode and a
$\lambda^{(1)}$ hole, giving rise to
\begin{equation}
\Delta = 2J - 2\int_0^\infty \frac{\e^{-\omega}}{1+\e^{-\omega}}
\d\omega = 2(J-\log 2).  
\end{equation}
This gap thus closes at $J_{\rm c,1}=\log 2$ and for
$J<J_{\rm c,1}$ there are three massless excitations and
four massive ones corresponding to the triplet states. These become
massless at a second critical point, $J_{\rm c,2}$, whose value can not
be calculated analytically. This point continuously connects to the
critical point $J_{\rm c}^+$ on the line $J'=J$.

\section{Phases}
\label{se:phases}
\begin{figure}[htb]
\centerline{\epsfig{file=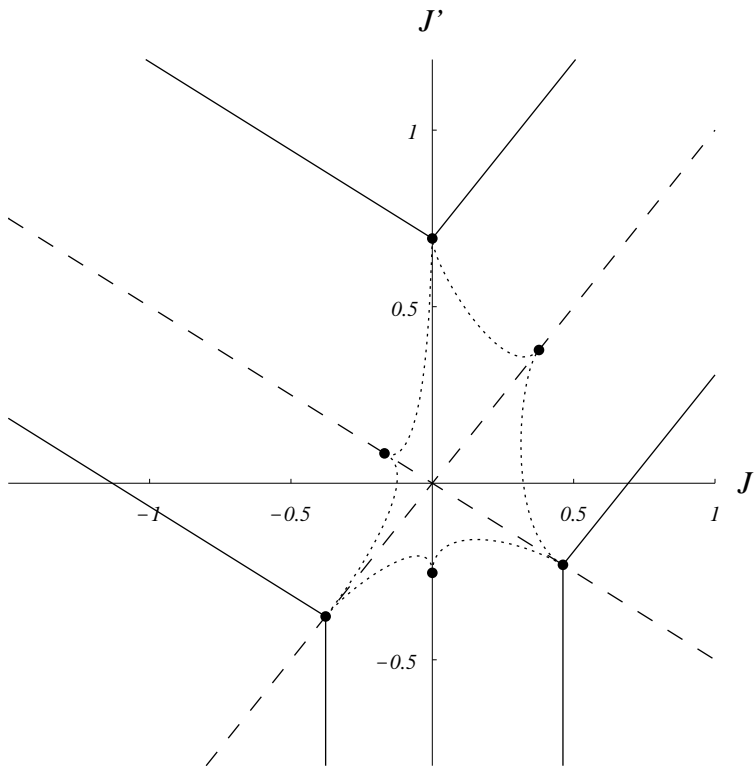}}
\vspace{20pt}
\caption{Phase diagram. The dashed lines are $J'=J$ and $2J'+J=0$
which together with the y-axis divide phase space into six
regions. The dotted lines are sketches of phase boundaries.}
\label{fig:phase}
\end{figure}

As already noted, the different choices of the reference state do not
change the physics, but each of them is convenient in some part of the
phase diagram. To appreciate this, it is helpful to divide the phase
diagram into six regions defined by the lines $J=0$, $J'=J$ and
$2J'+J=0$ as in Fig. \ref{fig:phase}. In each of these regions, the
corresponding choice of reference state of Section \ref{se:model} is
the most convenient.  

\begin{itemize}
\item[1.] $J>0$, $J-J'>0$ and $2J'+J>0$.

This part of the phase diagram contains the ladder for $J>0$. As in
Section \ref{se:ladder} we choose to describe the spectrum as in case
1. of the list in Section \ref{se:model}. For $J \gg J'$ it is obvious
that for the ground state one must have $M_2=0$. The ground state thus
is formed by the doublet $\{|0\rangle, |1\rangle \}$, which is
effectively a spin-$\frac{1}{2}$ XXX chain. The system is
critical with one massless excitation and six massive ones. The gap
can again be calculated by creating a $\lambda^{(2)}$ mode and a
$\lambda^{(1)}$ hole ($M_2 \rightarrow 1$, $M_1 \rightarrow M_1-1$) in
the Bethe Ansatz equations, and is given by
\begin{equation}
\Delta = 2(J-J') - 2\log 2.
\end{equation}
Thus there is a critical line on which this gap vanishes, given by
\begin{equation}
J'_{\rm c} = J - \log 2. 
\label{eq:2+2to2}
\end{equation}
Above this line some of the excitations on the
$\{|0\rangle,|1\rangle\}$ doublet become massless. We now
investigate which ones. From the form of the energy it follows that if
$2J'+J$ is large, $M_4=0$ and only the $\{|2\rangle,|3\rangle\}$ doublet
excitations have become massless. Thus for large $2J'+J$ and above the
critical line (\ref{eq:2+2to2}), the ground state is given by the two
spin-$\frac{1}{2}$ doublets and the quadruplet excitations are
massive. There is another phase transition where also the quadruplet
excitations become massless, i.e. $M_4\neq 0$ for the ground state, and
its location on the line (\ref{eq:2+2to2}) is given by
\begin{equation}
2J'_{\rm c}+J=0,\quad {\rm or}\quad (J,J') = (\case{2}{3}\log 2, 
-\case{1}{3}\log 2).
\end{equation}
Its location changes as $M_2$ increases, which can be calculated using
perturbative calculations (see Appendix \ref{ap:perturb}). It
ultimately reaches the value of 
\begin{equation}
(J,J') = (\case{1}{12}(\log 4 + \pi),\case{1}{12}(\log 4 + \pi)), 
\end{equation}
on the line $J'=J$ where three Fermi seas are completely filled
instead of just one. 

\item[2.] $J>0$, $J-J'<0$ and $2J'+J>0$.

By the same reasoning as above the ground state is again formed by a
doublet for $J'\gg J$, but now $\{|2\rangle, |3\rangle \}$ since we
use a different reference state. The system again is critical with one
massless excitation and six massive ones. The gap can be calculated
similarly and 
the critical line on which this gap vanishes is given by
\begin{equation}
J'_{\rm c} = J + \log 2. \label{eq:2pto2+2} 
\end{equation}
Below this line the $\{|0\rangle,|1\rangle\}$ doublet
excitations have become massless. There is another phase transition
where also the quadruplet excitations become massless, i.e. $M_4\neq
0$ for the ground state, and its location on the line
(\ref{eq:2pto2+2}) is given by 
\begin{equation}
(J,J') = (0,\log 2). 
\end{equation}
Its location changes as $M_2$ increases reaching the previously
calculated value,
\begin{equation}
(J,J') = (\case{1}{12}(\log 4 + \pi),\case{1}{12}(\log 4 + \pi)), 
\end{equation}
on the line $J'=J$ where three Fermi seas are completely filled
instead of just one. 

\item[3.] $J<0$, $J-J'<0$ and $2J'+J>0$. 

Here we find the phase boundary,
\begin{equation}
2J'_{\rm c}+J = 2\log 2,
\end{equation}
where the spin-$\frac{3}{2}$ quadruplet excitations on the
$\{|2\rangle,|3\rangle\}$ doublet ground state become massless. The
point on this line where also the spin-$\frac{1}{2}$ doublet
$\{|0\rangle,|1\rangle\}$ becomes massless is given by
\begin{equation}
J=0.
\end{equation}
As before, its location changes, reaching
\begin{equation}
(J,J') = (-\frac{\alpha_6}{3},\frac{\alpha_6}{6}),
\quad \alpha_6 \approx 0.5076...,
\end{equation}
on the line $2J'+J=0$.

\item[4.] $J<0$, $J-J'<0$ and $2J'+J<0$. 

In this region the ground state for large and negative $2J'+J$ is the
quadruplet. It's smallest excitations, those of the
$\{|2\rangle,|3\rangle\}$ doublet, become massless at
\begin{equation}
2J'+J = -\case{1}{4}(\pi + \log 4).
\end{equation}
On this line the other doublet becomes massless at 
\begin{equation}
(J,J') = (-\case{1}{12}(\pi + \log 4),-\case{1}{12}(\pi + \log 4)), 
\end{equation}
on the line $J'=J$. This point moves to 
\begin{equation}
(J,J') = (-\frac{\alpha_6}{3},\frac{\alpha_6}{6}), 
\end{equation}
on the line $2J'+J=0$.

\item[5.] $J<0$, $J-J'>0$ and $2J'+J<0$. 

In this region the quadruplet is still the ground state for large and
negative $J$, but now the lowest excitations are the
$\{|0\rangle,|1\rangle\}$ doublet. They become massless at 
\begin{equation}
J = -\case{1}{12}(\pi + \log 4), 
\end{equation}
independent of $J'$. The other doublet becoms massless at
\begin{equation}
(J,J') = (-\case{1}{12}(\pi + \log 4),-\case{1}{12}(\pi + \log 4)), 
\end{equation}
on the line $J'=J$ and this point extends to 
\begin{equation}
(J,J') = (0,-\frac{\alpha_6}{2}), 
\end{equation}
on the line $J=0$.

\item[6.] $J>0$, $J-J'>0$ and $2J'+J<0$. 

Here, the ground state for large $J$ is the $\{|0\rangle,|1\rangle\}$
doublet, and the smallest excitations are the quadruplet which become
massless at 
\begin{equation}
J = \case{2}{3} \log 2, 
\end{equation}
independent of $J'$. The other, doublet, excitations become massless
at 
\begin{equation}
(J,J') = (\case{2}{3} \log 2,-\case{1}{3} \log 2), 
\end{equation}
on the line $2J'+J=0$. On the other phase boundary, $J=0$, this point
is
\begin{equation}
(J,J') = (0,-\frac{\alpha_6}{2}).
\end{equation}
\end{itemize}

Our findings, along with the phase boundaries, are 
summarized in Fig. \ref{fig:phase}. 

\section{Conclusion}

In this paper we have calculated the phase diagram of the integrable
anisotropic 3-leg quantum spin tube defined in (\ref{eq:Ham}). The
model differs from the usual Heisenberg spin ladder in that it
includes additional four- and six-body terms in the leg
Hamiltonian (\ref{eq:hleg}). The phase diagram, as calculated from 
the Bethe Ansatz solution, is given in Fig. \ref{fig:phase}. The
straight line phase boundaries can be calculated exactly as well as
the critical points on the lines where the rung Hamiltonian becomes
partially degenerate. For the location of the dotted phase boundaries
one has to resort to perturbative calculations which show that they
end parallel to the degeneracy lines. In Fig. \ref{fig:phase} we have
sketched the full form of these phase boundaries. It is to be hoped
that their precise location may be accurately determined numerically. 

Since the model has an odd number of legs, it is critical in the
entire phase space and has no true gap. It is however the simplest
example with more than one quantum phase transition for $J>0$ and
serves as an example for what to expect for higher leg ladders. In
general, if the rung Hamiltonian has $n$ different eigenvalues, one
may expect $n-1$ different phase transitions along a generic line
between the phase with the lowest and highest total spin. At each of
these one expects to find quantum critical behaviour, which will however
diminish in magnitude. 

The 3-leg quantum spin tube is also integrable in the presence of
a magnetic field. The corresponding 3-leg Heisenberg model has been
shown to exhibit magnetization plateaus by means of renormalization
group calculations based on the bosonized effective continuum theory
\cite{COAIQ}. The exact calculation of the magnetic behaviour of the
present model will thus be of considerable interest. 

\acknowledgments
This work has been supported by the Australian Research Council.

\appendix
\section{Gap calculation}
\label{ap:gap}

As is well known, the Bethe Ansatz equations (\ref{eq:BAE}) may be
written as integral equations in the thermodynamic limit
$(L\rightarrow \infty)$,
\begin{equation}
a_1(\lambda) = \rho^{(1)}(\lambda) + \int_{-\lambda^{(1)}_{\rm
F}}^{\lambda^{(1)}_{\rm F}} a_2(\lambda-\lambda') \rho^{(1)}(\lambda')
\d\lambda' - \int_{-\lambda^{(2)}_{\rm F}}^{\lambda^{(2)}_{\rm F}}
a_1(\lambda-\lambda') \rho^{(2)}(\lambda') \d\lambda',
\label{eq:BAEtherm1}
\end{equation}
\begin{eqnarray}
\rho^{(r)}(\lambda) &+& \int_{-\lambda^{(r)}_{\rm
F}}^{\lambda^{(r)}_{\rm F}} a_2(\lambda-\lambda') \rho^{(r)}(\lambda')
\d\lambda' = \nonumber\\ 
&&\int_{-\lambda^{(r-1)}_{\rm F}}^{\lambda^{(r-1)}_{\rm F}}
a_1(\lambda-\lambda') \rho^{(r-1)} \d\lambda' +
\int_{-\lambda^{(r+1)}_{\rm F}}^{\lambda^{(r+1)}_{\rm F}}
a_1(\lambda-\lambda') \rho^{(r+1)}(\lambda') \d\lambda',
\label{eq:BAEtherm2}  
\end{eqnarray} 
where $\rho^{(r)}$ are the distributions of Bethe Ansatz roots and
holes and
\begin{equation}
a_n(\lambda) = \frac{n}{2\pi} \frac{1}{\lambda^2
+ (\frac{n}{2})^2}. 
\end{equation}
In the sequel we will make use of the following convention
for the Fourier transform,
\begin{equation}
\hat{f}(\omega) = \frac{1}{2\pi} \int_{-\infty}^{\infty} f(k)
\e^{-\i k\omega} \d k,\quad
f(k) = \int_{-\infty}^{\infty} \hat{f}(\omega) \e^{\i k\omega} \d
\omega,
\end{equation}
and of the results,
\begin{eqnarray}
\int_{-\infty}^{\infty} a_n(k) \e^{-\i k\omega} \d k &=&
\e^{-\frac{n}{2} |\omega|},\quad {\rm for}\; n>0,\\
\int_{0}^{\infty} \frac{\e^{-a \omega}}{1+\e^{-\omega}} \d\omega &=&
\frac{1}{2} \left( \psi^{(0)}(\case{a+1}{2}) -
\psi^{(0)}(\case{a}{2}) \right),\quad \psi^{(0)}(z) 
= \frac{\Gamma'(z)}{\Gamma(z)}.
\end{eqnarray}

As an example we give here the calculation of the gap for the tube
when $J'=J >0$. The ground state for $J \gg 0$ consists of three
completely filled Fermi seas. The Bethe Ansatz equations
(\ref{eq:BAEtherm1}) and (\ref{eq:BAEtherm2}) may therefore be solved
with Fourier transforms. A quadruplet excitation is created with a
$\lambda^{(4)}$ mode and a $\lambda^{(3)}$ hole. Following Wang
\cite{W}, we will denote the changes in the distribution functions as
a consequence of this particle-hole pair creation by
$\delta\rho^{(r)}$. The Fourier transformed Bethe Ansatz equations
then become (with some abuse of notation)    
\begin{eqnarray}
0 &=& \delta\hat{\rho}^{(1)}(\omega) (1+\e^{-|\omega|}) -
\delta\hat{\rho}^{(2)}(\omega) \e^{-\frac{1}{2}|\omega|} \nonumber\\ 
\delta\hat{\rho}^{(2)}(\omega) (1+\e^{-|\omega|}) &=&
\delta\hat{\rho}^{(1)}(\omega) \e^{-\frac{1}{2}|\omega|} +
\delta\hat{\rho}^{(3)}(\omega) \e^{-\frac{1}{2}|\omega|} -
\frac{1}{2\pi L} \e^{-\i \lambda^{(3)}\omega-\frac{1}{2}|\omega|}
\nonumber \\
\delta\hat{\rho}^{(3)}(\omega)(1+\e^{-|\omega|}) &=& \frac{1}{2\pi L}
\e^{-\i \lambda^{(3)}\omega-|\omega|} + \delta\hat{\rho}^{(2)}(\omega)
\e^{-\frac{1}{2}|\omega|} + \frac{1}{2\pi L} \e^{-\i
\lambda^{(4)}\omega-\frac{1}{2} |\omega|}\nonumber,
\end{eqnarray}
from which we obtain 
\begin{equation}
\delta\hat{\rho}^{(1)}(\omega) = \frac{1}{2\pi L} \frac{ \e^{-\i
\lambda^{(4)} \omega} - \e^{\frac{1}{2}|\omega| -\i \lambda^{(3)}\omega}}
{4\cosh \omega \cosh \frac{\omega}{2}}.
\end{equation}
Thus it follows that the smallest gap ($\lambda^{(3)} \rightarrow
\infty$, $\lambda^{(4)} \rightarrow 0$) is equal to
\begin{equation}
\Delta = 3J - 2\pi L \int_{-\infty}^{\infty} \e^{-\frac{1}{2} |\omega|}
\delta\hat{\rho}^{(1)}(\omega) \d \omega = 3J - \case{1}{4}(\log 4 + \pi).
\end{equation}

\section{Perturbative calculations}
\label{ap:perturb}

Suppose we are in region I in the phase where both doublet excitations are
massless, but close to the phase boundary $J'=J-\log 2$. The
$\{|2\rangle,|3\rangle\}$ excitations have just become massless and
$M_2/L$ and $M_3/L$ have a small but finite value. The $\lambda^{(2)}$
Fermi sea will be filled up to some finite momentum $\Lambda$ while
those of $\lambda^{(1)}$ and $\lambda^{(3)}$ are completely
filled. By creating a $\lambda^{(4)}$ excitation, we change $\Lambda$
and the distribution functions. The Bethe Ansatz equations for this
situation are given by
\begin{eqnarray}
&&\rho^{(1)}(\lambda) + \int_{-\infty}^{+\infty}
a_2(\lambda-\lambda') \rho^{(1)}(\lambda') \d\lambda' = a_1(\lambda)  
+ \int_{-\tilde{\Lambda}}^{+\tilde{\Lambda}} a_1(\lambda-\lambda')
\rho^{(2)}(\lambda') \d\lambda', \label{eq:BApert1}\\
&&\rho^{(2)}(\lambda) + \int_{-\tilde{\Lambda}}^{+\tilde{\Lambda}}
a_2(\lambda-\lambda') \rho^{(2)}(\lambda') \d\lambda' = 
\int_{-\infty}^{+\infty} a_1(\lambda-\lambda')
\rho^{(1)} (\lambda') \d\lambda' + \int_{-\infty}^{+\infty}
a_1(\lambda-\lambda') \rho^{(3)}(\lambda') \d\lambda',
\label{eq:BApert2} \\
&&\rho^{(3)}(\lambda) + \int_{-\infty}^{+\infty}
a_2(\lambda-\lambda') \rho^{(3)}(\lambda') \d\lambda' =
\int_{-\tilde{\Lambda}}^{+\tilde{\Lambda}} a_1(\lambda-\lambda')
\rho^{(2)} (\lambda') \d\lambda' + \frac{a_1(\lambda)}{L},
\label{eq:BApert3}  
\end{eqnarray} 
with the conditions
\begin{equation}
\int_{-\tilde{\Lambda}}^{+\tilde{\Lambda}} \rho^{(2)}(\lambda) \d
\lambda = \frac{M_2}{L},\quad \int_{-\infty}^{+\infty}
\rho^{(3)}(\lambda) \d \lambda = \frac{M_3}{L}. \label{eq:dens}
\end{equation}
These equations can be solved perturbatively in $\tilde{\Lambda}$ which we
will do up to $O(\tilde{\Lambda}^3)$. With the help of
\begin{eqnarray}
\int_{-\tilde{\Lambda}}^{+\tilde{\Lambda}} a_n(\lambda-\lambda')
\rho^{(2)}(\lambda') \d\lambda' &=& 2\tilde{\Lambda} a_n(\lambda) \rho^{(2)}
(0) +\case{1}{3} \tilde{\Lambda}^3 (a_n''(\lambda) \rho^{(2)} (0) +
a_n(\lambda) \rho^{(2)}{}''(0) ), \nonumber\\ 
&=& a_n(\lambda) \frac{M_2}{L} + \case{1}{3} \tilde{\Lambda}^3 a_n''(\lambda)
\rho^{(2)}(0),
\end{eqnarray}
we find
\begin{eqnarray}
\hat{\rho}^{(1)}(\omega) &=& \frac{1}{2\pi} \frac{1}{2\cosh
\frac{\omega}{2}} \left( 1 + \frac{M_2}{L} - \frac{\log 2}{3\pi}
\tilde{\Lambda}^3 \omega^2 (1+\frac{1}{L}) \right),\\
\rho^{(2)}(0) &=& \frac{\log 2}{\pi}(1+\frac{1}{L}) +
O(\tilde{\Lambda}). \label{eq:rho2nil}
\end{eqnarray}
Since we want to keep $M_2$ constant it follows from
(\ref{eq:rho2nil}) that
\begin{equation}
\tilde{\Lambda} = \Lambda(1-\frac{1}{L}) + O \left(\frac{1}{L^2} \right). 
\end{equation}
Finally, the energy is given by
\begin{eqnarray}
E &=& -2\pi L \int_{-\infty}^{+\infty} \hat{\rho}^{(1)}(\omega)
\e^{-\frac{1}{2}|\omega|}\d \omega + 2(J-J')M_2 + 2J'+J\nonumber\\
&=& - 2L \log2(1+\frac{M_2}{L}) + L \tilde{\Lambda}^3 \frac{\zeta(3)
\log 2}{\pi} (1+\frac{1}{L}) + 2(J-J')M_2 + 2J'+J\nonumber \\
&=& -2L\log2 - 2M_2\left[ \log2 - \case{1}{4}\zeta(3) \Lambda^2  -J +
J'\right] +2J'+J - \Lambda^3 \zeta(3) \frac{\log 2}{\pi} + O(\frac{1}{L}).  
\end{eqnarray}
It thus follows that the Fermi momentum $\Lambda$ is defined by
\begin{equation}
J' = J - \log2 + \case{1}{4}\zeta(3) \Lambda^2,
\end{equation}
and that the $\lambda^{(4)}$ become massless on this line at the point
\begin{equation}
2J'+J = \Lambda^3 \zeta(3) \frac{\log 2}{\pi}.
\end{equation}
We thus conclude that this phase boundary ends in the point 
\begin{equation}
(J,J') = (\case{2}{3} \log 2,-\case{1}{3} \log 2)
\end{equation}
parallel to the line $2J'+J = 0$.


\begin{references}

\bibitem{DR} See, e.g., E. Dagotto and T. M. Rice, 
Science {\bf 271} 618; E. Dagotto, cond-mat/9908250, and references
therein. 

\bibitem{A} M. Azuma et al., Phys. Rev. Lett. {\bf 73}, 3463 (1994).

\bibitem{C} G. Chaboussant et al., Phys. Rev. B {\bf 55}, 3046 (1997).

\bibitem{FR} H. Frahm and C. R\"odenbeck, Europhys. Lett. 
{\bf 33}, 47 (1996); J. Phys. A {\bf 30}, 4467 (1997). 

\bibitem{AFW} S. Albeverio, S.-M. Fei and Y. Wang, Europhys. Lett. 
{\bf 47}, 364 (1999).

\bibitem{BMb} M. T. Batchelor and M. Maslen, cond-mat/9907480. 

\bibitem{W} Y. Wang, Phys. Rev. B {\bf 60}, 9236 (1999). 

\bibitem{BMa} M. T. Batchelor and M. Maslen, 
J. Phys. A {\bf 32}, L377 (1999). 

\bibitem{F} H. Fan, cond-mat/9908028.

\bibitem{FK} H. Frahm and A. Kundu, cond-mat/9910104.

\bibitem{BGLM} M. T. Batchelor, J. de Gier, J. Links and M. Maslen,
cond-mat/9911043.

\bibitem{mp} See, e.g., A.K. Kolezhuk and H.-J. Mikeska,
Int. J. Mod. Phys. B {\bf  12}, 2325 (1998), and references therein. 

\bibitem{S} B. Sutherland, Phys. Rev. B {\bf 12}, 3795 (1975).

\bibitem{COAIQ} R. Citro, E. Orignac, N. Andrei, C. Itoi and S. Qin, 
cond-mat/9904371.

\end{references}
\end{document}